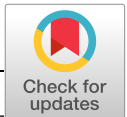

## CONDENSED MATTER PHYSICS

# Potential Lifshitz transition at optimal substitution in nematic pnictide Ba$_{1−x}$Sr$_x$Ni$_2$As$_2$

Dushyant M. Narayan[1]*, Peipei Hao[1], Rafał Kurleto[1], Bryan S. Berggren[1], A. Garrison Linn[1], Christopher Eckberg[2], Prathum Saraf[2], John Collini[2], Peter Zavalij[3], Makoto Hashimoto[4], Donghui Lu[4], Rafael M. Fernandes[5], Johnpierre Paglione[2,6], Daniel S. Dessau[1]*

BaNi$_2$As$_2$ is a structural analog of the pnictide superconductor BaFe$_2$As$_2$, which, like the iron-based superconductors, hosts a variety of ordered phases including charge density waves (CDWs), electronic nematicity, and superconductivity. Upon isovalent Sr substitution on the Ba site, the charge and nematic orders are suppressed, followed by a sixfold enhancement of the superconducting transition temperature ($T_c$). To understand the mechanisms responsible for enhancement of $T_c$, we present high-resolution angle-resolved photoemission spectroscopy (ARPES) measurements of the Ba$_{1−x}$Sr$_x$Ni$_2$As$_2$ series, which agree well with our density functional theory (DFT) calculations throughout the substitution range. Analysis of our ARPES-validated DFT results indicates a Lifshitz transition and reasonably nested electron and hole Fermi pockets near optimal substitution where $T_c$ is maximum. These nested pockets host Ni $d_{xz}/d_{yz}$ orbital compositions, which we associate with the enhancement of nematic fluctuations, revealing unexpected connections to the iron-pnictide superconductors. This gives credence to a scenario in which nematic fluctuations drive an enhanced $T_c$.

## INTRODUCTION

The interplay between normal-state electronically ordered states and superconductivity has long been identified as key to understanding high-temperature superconductivity. In the case of the iron pnictide superconductors, such as BaFe$_2$As$_2$, superconductivity coexists with spin-density wave order and electronic nematicity [for a recent review, see (1)].

Recently, BaNi$_2$As$_2$, a structural analog of BaFe$_2$As$_2$, crystallizing in the same high-temperature body-centered tetragonal (BCT) structure (space group $I4/mmm$) (2, 3), has been found to exhibit charge density wave (CDW) order and electronic nematicity without any evidence for magnetic order (4–11).

As a function of Sr substitution on the Ba site, a rich phase diagram was uncovered (see Fig. 1A), exhibiting bidirectional incommensurate charge order (12) and electronic nematic order (5), followed by a structural transition into a low-symmetry triclinic phase (space group $P1$), and a commensurate CDW lock-in transition at lower temperatures. These phases' corresponding transition temperatures are suppressed upon increasing Sr substitution, such that at a critical Sr concentration of $x \sim 0.72$, they are fully suppressed, and the superconducting transition temperature is increased sixfold (5).

The resulting phase diagram (see Fig. 1A) of Ba$_{1−x}$Sr$_x$Ni$_2$As$_2$ thus shows remarkable similarity to the prototypical iron-pnictide superconductor BaFe$_2$As$_2$, displaying a putative nematic quantum critical point (QCP) (13), but with the notable absence of magnetic ordering. Several theoretical works have shown that nematic fluctuations can lead to a sharp enhancement of $T_c$ (14–19), even if another primary pairing mechanism is at play. In iron pnictides such as BaFe$_2$As$_2$, it is difficult to disentangle contributions from spin and magnetic fluctuations toward pairing. In iron chalcogenides such as S-doped FeSe, where magnetic order is absent, the putative nematic QCP seems to have little impact on $T_c$ (20). Therefore, BaNi$_2$As$_2$ offers a unique platform to investigate the interplay between electronic nematicity and pairing (21), without the interfering effects of magnetic fluctuations (9).

In this work, we resolve the electronic structure of the high-temperature tetragonal phase of Ba$_{1−x}$Sr$_x$Ni$_2$As$_2$ as a function of Sr substitution, through a combination of density functional theory (DFT) and angle-resolved photoemission spectroscopy (ARPES) experiments. Some of the early ARPES work on the parent compound BaNi$_2$As$_2$ found reasonable agreement with DFT (22) but did not see evidence of charge ordering and band back folding. More recent work found evidence of the unidirectional CDW in the triclinic phase of the parent ($x = 0$) compound by performing ARPES under strain (6). In contrast, our work reports successful measurements of the Sr-substituted compounds as well as accurate DFT calculations that use structural data refined from x-ray diffraction (XRD).

Unlike the case of the iron pnictides, for this nickel pnictide, we find very good agreement with DFT, with negligible chemical potential shifts and minimal band mass enhancements. After validating the DFT results with the ARPES data along the main high symmetry planes, we analyze the DFT band structure at other $k_z$ values to elucidate the evolution of the electronic structure as a function of Sr substitution. We identify a potential Lifshitz transition near optimal substitution in which the N-point hole pockets sink below the Fermi level (see the Brillouin zone in Fig. 1B). Preceding the Lifshitz transition, the Ni $d$-orbital content of the innermost P-point electron pocket, which is displaced from the N points by the momenta (π,0,0) and (0,π,0), changes from $d_{x^2−y^2}$ dominated at low Sr concentration to $d_{xz}/d_{yz}$ dominated near the Lifshitz transition. The N pocket, on the other hand, has mixed $d_{xz}/d_{yz}$ and $d_{xy}$ orbital character, with increasing dominance of the $d_{xz}/d_{yz}$ orbitals near the Lifshitz transition. The similar $d_{xz}/d_{yz}$ orbital compositions between the hole and electron pockets set the conditions for the enhancement of $B_{1g}$ nematic fluctuations, akin to the situation involving the $\Gamma$ and X Fermi pockets in the iron pnictides—the latter being separated by $(\pm\pi,\pi,0)$, leading to $B_{2g}$ nematicity (23). The fact that the substitution level at which this proposed Lifshitz transition takes place corresponds to the substitution level where superconductivity is maximal and nematic fluctuations are enhanced is consistent with the idea that the pairing interaction acquires important contributions from fluctuations of the nematic order near the putative QCP.

[1]Center for Experiments on Quantum Materials, Department of Physics, University of Colorado, Boulder, CO 80309, USA. [2]Maryland Quantum Materials Center, Department of Physics, University of Maryland, College Park, MD 20742, USA. [3]Department of Chemistry, University of Maryland, College Park, MD 20742, USA. [4]Stanford Synchrotron Radiation Lightsource, SLAC National Accelerator Laboratory, Menlo Park, CA 94025, USA. [5]School of Physics and Astronomy, University of Minnesota, Minneapolis, MN 55455, USA. [6]Canadian Institute for Advanced Research, Toronto, ON M5G-1Z8, Canada.
*Corresponding author. Email: duna1846@colorado.edu (D.M.N.) dan.dessau@colorado.edu (D.S.D.)

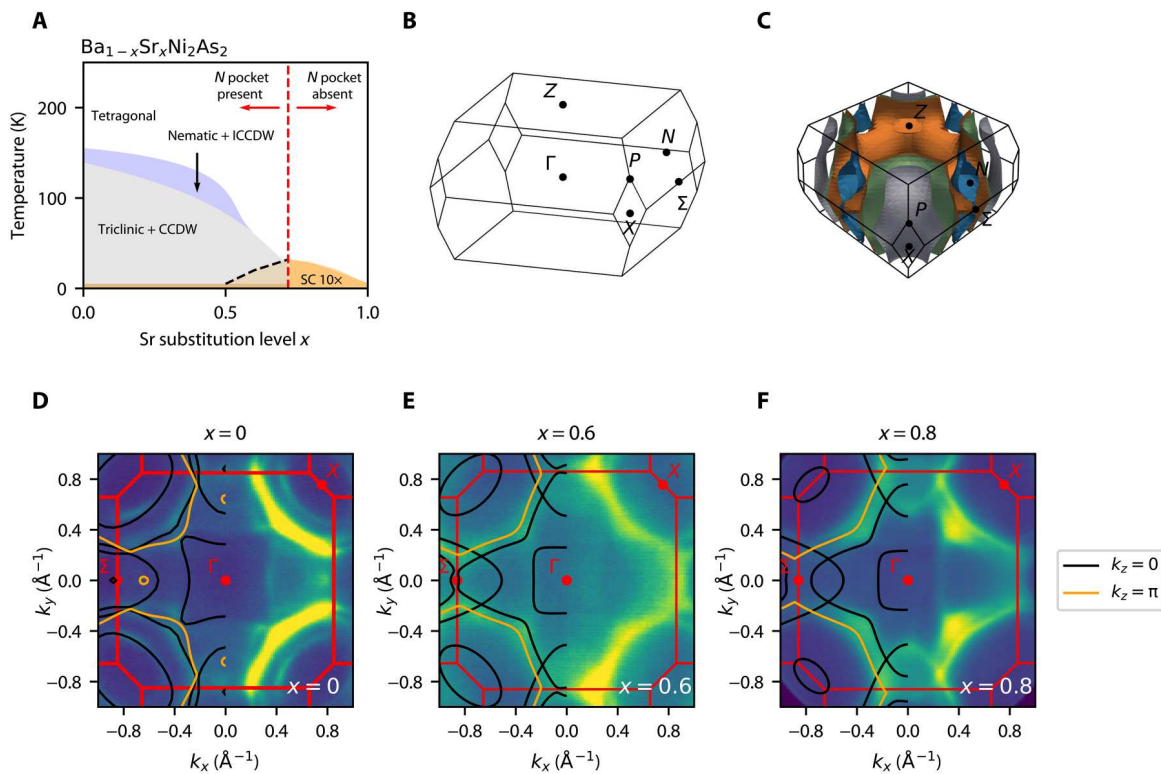

**Fig. 1. Schematic phase diagram of $Ba_{1-x}Sr_xNi_2As_2$ and ARPES Fermi Surface (FS) maps as a function of Sr substitution with results from density functional theory (DFT) calculations overlaid.** (**A**) Temperature and Sr substitution dependence of the triclinic/commensurate charge density wave (CCDW) phase transitions, as well as of the nematic/incommensurate charge density wave (ICCDW) phase transitions. The superconducting (SC) phase is shown with $T_c$ multiplied by 10, highlighting the sixfold enhancement at a critical substitution of approximately $x = 0.72$, as reported in (5). For $x$ in the range of $0.5 < x < 0.72$, the SC appears to be filamentary (12). Red dashed line indicates the substitution level of the Lifshitz transition in the tetragonal phase, where the N = $(\pi,0,\pi/2)$–point hole pocket [blue pocket in (C)] is pushed below the Fermi level. (**B**) BCT Brillouin zone and its high-symmetry points. (**C**) 3D rendering of $x = 0$ parent compound FS, as obtained from our DFT calculations.. (**D** to **F**) ARPES FS maps of three substitution levels, all taken in the tetragonal phase at 200 K and with photon energies of 100, 115, and 75 eV, respectively. The $k_z$ values were determined from photon-energy scans to be cutting through the $\Gamma$ point. Each FS plot also contains overlays of the FS obtained from DFT by using each respective structure determined by XRD. Because of either an extrinsic scattering process or surface states, additional sets of replica bands that appear to be from $k_z = \pi$ are observed, as shown by the orange replica lines.

### RESULTS

Figure 1B shows the three-dimensional (3D) Brillouin zone with the high-symmetry points labeled, while Fig. 1C shows our DFT-calculated Fermi surface (FS) for the parent ($x = 0$) compound $BaNi_2As_2$. Two of the FS sheets centered at the X = $(\pi,\pi,0)$ point are approximately cylindrical with minimal $k_z$ dependence, as expected for a layered material. A third FS sheet centered at the X point also has a piece enclosed in $k_z$, which can be seen midzone along the $\Sigma - \Gamma - \Sigma$ direction. Note also the small pocket shown in blue that is centered at the N = $(\pi,0,\pi/2)$ point [and another one at the symmetry-related $(0,\pi,\pi/2)$ point], which will undergo the Lifshitz transition as a function of Sr substitution. In Fig. 1 (D to F), 2D ARPES FS maps of the $x = 0$, $x = 0.6$, and $x = 0.8$ compounds are shown with the DFT FS overlaid. Photon energies were chosen to place the data in the $k_z = 0$ plane (black lines) for each respective compound. Because of the changing lattice parameters as a function of Sr substitution, different photon energies are required to reach $k_z = 0$ for each compound (see the Supplementary Materials). In addition, the data show evidence for states scattered from the $k_z = \pi$ plane (orange lines), which could be due to an extrinsic process that scatters states along $k_z$, or could be evidence of surface states.

The agreement between the DFT calculations and the ARPES data is overall good to excellent for all substitution levels studied. The FS at the $k_z = 0$ plane for all compounds consists of three main electron pockets, two centered at the X point, and a third that has a piece of FS enclosed around the X point and another piece that is enclosed in $k_z$ midzone, and can be seen in the Γ to Σ direction. There is also one hole pocket at the N point, shown in the 3D FS of Fig. 1C, which is not shown in the ARPES FS maps, because it is located in the $k_z = \pi/2$ plane. Upon Sr substitution, the innermost electron pocket centered at the X point is markedly reduced in size. Because Sr is isovalent to Ba, this process should be charge neutral, implying that the charge must be redistributed between the pockets as a function of Sr substitution. We will show later that the shrinking of the X-point electron pocket is linked to the Lifshitz transition of the hole pocket at the N point, for which other $k_z$ values need to be explored.

In Fig. 2, ARPES energy dispersion $E(k)$ cuts of the $x = 0$, $x = 0.6$, and $x = 0.8$ compounds are shown, together with the corresponding DFT results. The energy dispersion cuts also show excellent agreement with DFT, similar to the FS maps, which, unlike in the iron pnictides, do not require shifting of the chemical potential or of the bottom/top of the bands (24, 25). The comparison to the DFT results reveals the existence in the spectra of states scattered from different $k_z$ planes, as indicated in the legend of Fig. 2. For the Σ − Γ − Σ cuts shown in Fig. 2 (B to D), an extra band originally centered at the Z point is seen at Γ (orange) for the $x = 0$ and $x = 0.6$ compounds. For the X-X cuts, one additional band from $k_z = \pi/2$ is also visible (in orange). To match the ARPES and DFT bands, the outer electron pockets required a small renormalization factor between 1.25 and 1.3, which indicates that the measured bands have slightly higher mass than those predicted by the DFT. This renormalization was performed about the Fermi level $E_f$, indicating that this is a self-energy effect, rather than a purely structural effect. This mass renormalization is much smaller than what is found in iron-based superconductors (1, 26), indicating the weak nature of correlations in the Ba$_{1-x}$Sr$_x$Ni$_2$As$_2$ family of materials.

Figure 3 shows ARPES FSs in the $k_z − k_x$ plane. Figure 3 (A and B) again shows the good agreement between the DFT-derived FSs and the ARPES spectra. In Fig. 3 (C,D and E), we can see an alternative view of the evolution of the FS as a function of Sr substitution. In particular, we can see the N pocket, highlighted in blue, "pinching off" in the $k_z$ direction near optimal substitution of $x \sim 0.75$. In Fig. 3A, if we interrogate the ARPES spectra closely, we can see that the spectral weight appears to be smooth and uninterrupted along the N-N direction in $k_z$, while in Fig. 3B, we can see that the spectral weight appears "interrupted" at the N point. We do not, however, in this case, see a direct view of the N pocket. The inability to directly view the N pocket in $k_z$ data in this way is related to the insensitivity of ARPES to states in the midzone in $k_z$. First, ARPES $k_z$ data suffer from reduced $k_z$ resolution due to short inelastic mean-free paths of photoelectrons emitted in the photoemission process (which leads to the surface sensitivity of ARPES spectra). In addition, ARPES has the greatest sensitivity to the states at the Brillouin zone edges ($k_z = 0$, $k_z = \pi$), where the bands are flat in $k_z$, giving them an enhanced density of states and greater cross section. Because of these two sets of issues, midzone states are harder to decouple from other states found in close proximity in $k_z$. Although this is not a direct method of resolving the N pocket, we believe that this signature of "gapping" of the spectral weight in the $k_z$ data lends credence to our proposed scenario of a Lifshitz transition occurring near the optimal substitution level of $x \sim 0.75$.

## DISCUSSION
Because of the excellent agreement between DFT and ARPES spectra as shown in Figs. 1 to 3, we now interrogate the DFT-derived electronic structure at other $k_z$ planes to look for systematic changes as a function of Sr substitution and the accompanying structural changes in the lattice. The excellent agreement also gives us additional confidence in the orbital projections predicted by DFT. For the electronic structure in the $k_z$ = 0 plane, these orbital projections, restricted here to the subspace of the Ni $d$-orbitals, indicate that the outer pockets at the X point have substantial $d_{xy}$ character (see the Supplementary Materials). Because these states are the ones that require a slight mass renormalization, as discussed in the previous section, this indicates that the $d_{xy}$ orbital is slightly more correlated than the other $d$-orbitals, a trend also seen in the iron-based superconductors (*1*, *26*). While the larger mass renormalization of $d_{xy}$ states points toward similar orbital-dependent physics as found in the iron-based superconductors, the relative weakness of the renormalization indicates weaker correlation effects in this material system. This is not unexpected, because Ni in BaNi$_2$As$_2$ has a nominal 3$d^8$ valence, while Fe has a nominal 3$d^6$ valence in BaFe$_2$As$_2$. Thus, BaNi$_2$As$_2$ is further away from the 3$d^5$ valence where correlations are expected to be the most important (*27*).

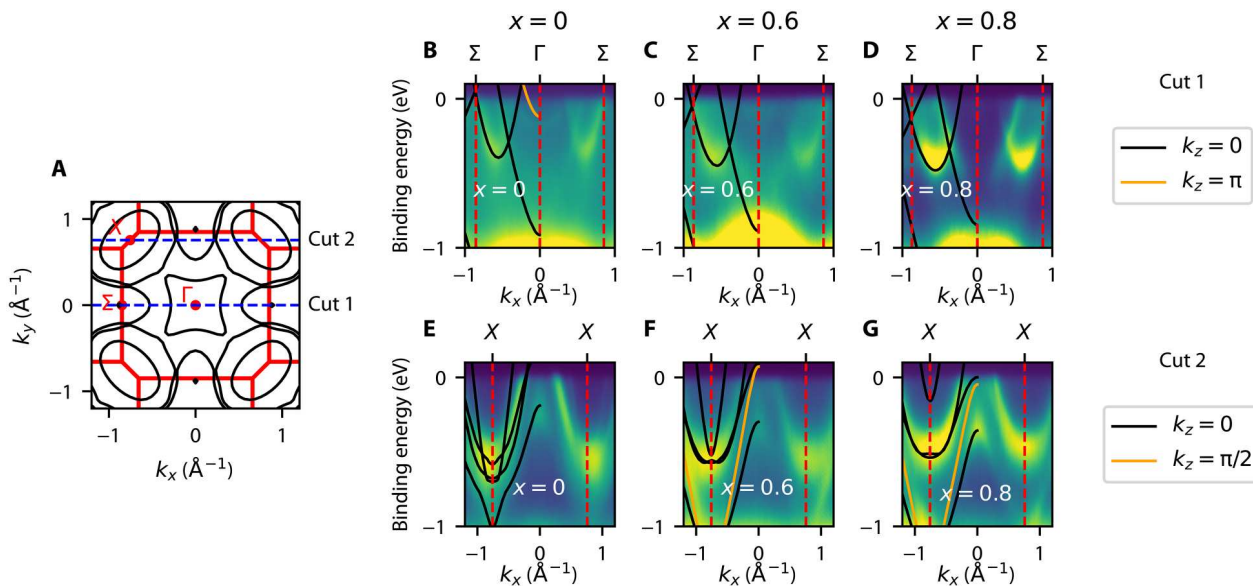

**Fig. 2. ARPES $E(k)$ energy-dispersion cuts as a function of Sr substitution.** (**A**) Schematic FS of the $x = 0$ compound with the Brillouin zone and high-symmetry points shown in red. Blue dashed lines indicate the k-space orientation of the cuts Σ − Γ − Σ (cut 1) and X-X (cut 2). (**B** to **D**) $E(k)$ energy dispersion along cut 1 for the $x = 0$, $x = 0.6$, and $x = 0.8$ compounds, respectively. Photon energies used were 100, 115, and 75 eV for $x = 0$, $x = 0.6$, and $x = 0.8$ compounds. Black and orange curves are the $E(k)$ dispersions obtained from DFT calculations with no renormalization or chemical potential shift but from different $k_z$ planes. (**E** to **G**) $E(k)$ energy dispersion along cut 2 for the $x = 0$, $x = 0.6$, and $x = 0.8$ compounds, respectively. Overlays in black/orange are also obtained from DFT calculations with no chemical potential shift but from different $k_z$ planes. The two bands that comprise the outermost of the two pockets centered at the X point required mass renormalizations about $E_f$ of 1.25, 1.25, and 1.3 for $x = 0$, $x = 0.6$, and $x = 0.8$, respectively. These bands have a dominant $d_{xy}$ character according to the DFT calculations (see the Supplementary Materials).

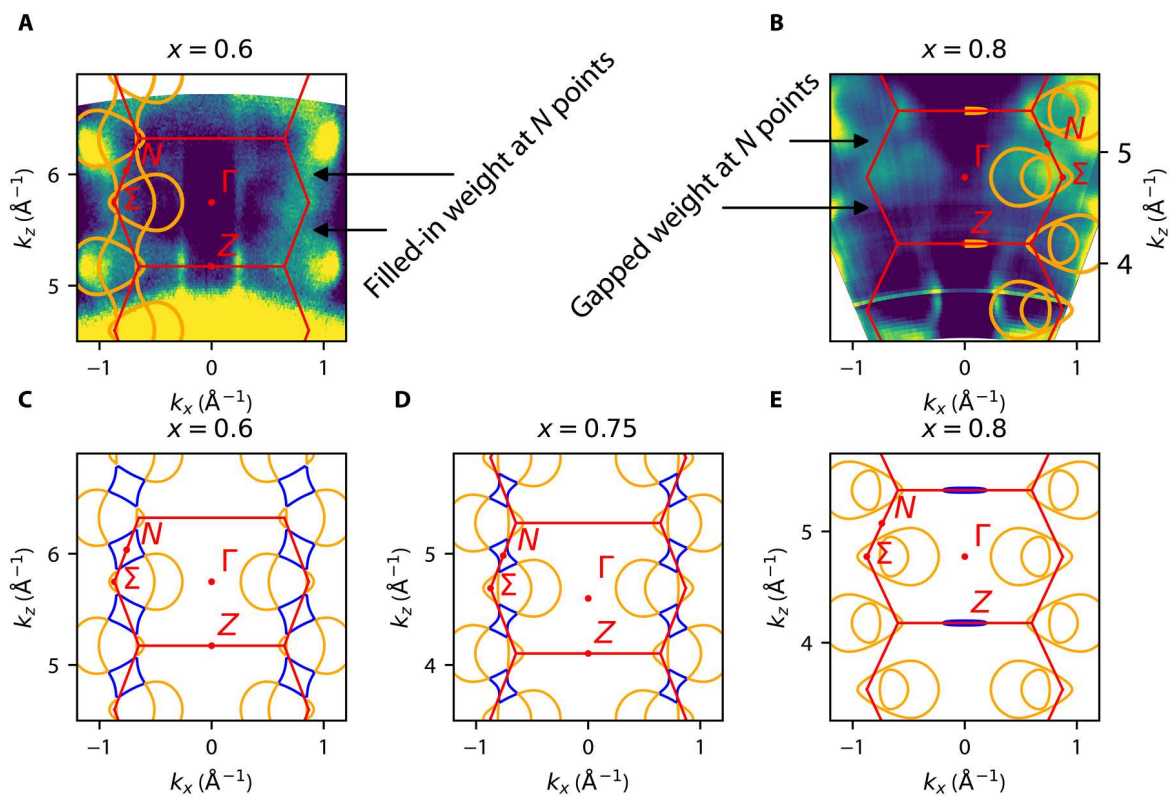

**Fig. 3. ARPES $k_z$ scans and DFT comparison.** (**A**) FS map of the $k_x$ – $k_z$ plane extracted from photon-energy scan of the $x$ = 0.6 compound with DFT overlay (orange). Scan shows continuous spectral weight along the N-N direction, indicating a possible signature of the N pocket. (**B**) FS map of the $k_x$ – $k_z$ plane extracted from photon-energy scan of the $x$ = 0.8 compound with DFT overlay (orange). In contrast to the $x$ = 0.6 scan, the spectral weight at the N points appears interrupted, potentially indicating the absence of the pocket at the Fermi level. (**C**) DFT-calculated FS in the $k_x$ – $k_z$ plane of the $x$ = 0.6 compound. The N pocket is highlighted in blue. (**D**) DFT-calculated FS in the $k_x$ – $k_z$ plane of the $x$ = 0.75 compound showing the N pocket "pinching off" (blue). (**E**) DFT-calculated FS in the $k_x$ – $k_z$ plane of the $x$ = 0.8 compound showing the completion of the Lifshitz transition. The pocket has completely disappeared, and the spectral weight is discontinuous along the N-N direction in $k_z$.

Figure 4 shows the FSs and Ni $d$-orbital projections as a function of Sr substitution in the N-P (i.e., $k_z$ = π/2) plane, where we recall, N = (π,0,π/2) [with a symmetry-related point at (0,π,π/2), and P = (π,π, π/2)]. Note that, as explained in the Supplementary Materials, orbitals from the As and even the Ba atoms also contribute substantial spectral weights for the Fermi pockets at this $k_z$ plane, particularly in the case of the innermost P pocket, for which the As contribution is slightly larger than the Ni contribution. For the purposes of our discussion, we will focus hereafter only on the Ni $d$-orbitals. At the Fermi level, the system has substantial $d_{xy}$ (red), $d_{x^2 - y^2}$ (orange), and $d_{yz}/d_{xz}$ character (blue/green). All Ni $d$-orbital projections show strong energy, and momentum dependence, as well as substantial hybridization. An important set of states to consider are those that form a small hole pocket at the N point (blue pocket in Fig. 1C). While this pocket is seen clearly in Fig. 4 (A to C), it is absent in Fig. 4D. As shown in Fig. 4E, which displays energy dispersion cuts along the P-N-P direction, the band giving rise to the N = (π,0,π/2)–point hole pocket is dominantly $d_{x^2 - y^2}$ in character at deeper binding energies, whereas it becomes predominantly $d_{yz}$ in character at the Fermi level—within the Ni $d$-orbital subspace, as explained above. Accordingly, the dominant Ni $d$-orbital for the symmetry-related Fermi pocket at N = (0,π,π/2) has a predominant $d_{xz}$ character at the Fermi level. Along the $N - \overline{\Gamma} - N$ direction, as shown by the energy cut of Fig. 4I, the hole-pocket bands have a dominant $d_{xy}$ character at deeper binding energies, which switches to a dominant $d_{yz}/d_{xz}$ character at the top of the bands, near the Fermi level.

Besides the orbital composition of the set of states at N and P, Fig. 4 also shows their evolution as a function of Sr substitution.

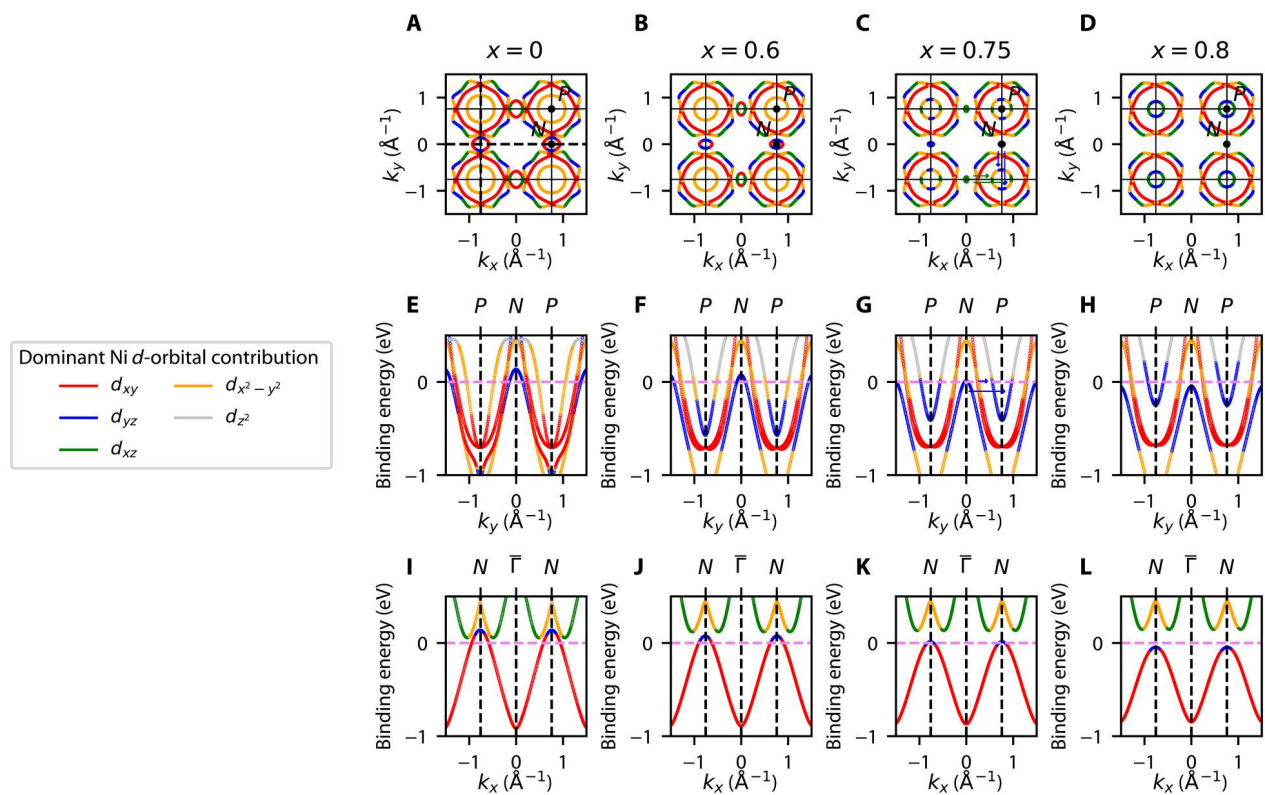

**Fig. 4. Calculated DFT FSs showing a Lifshitz transition as a function of Sr substitution.** (**A**) The FS of the $x = 0$ parent compound in the $k_z = \pi/2$ N-P plane. There are three electron pockets centered at the P point and one hole pocket centered at the N point. Vertical black dashed lines denote the P-N-P cut in (**E**) to (**H**), while horizontal dashed lines indicate the $N - \bar{\Gamma} - N$ cut in (**I**) to (**L**). The color scaling indicates the dominant orbital character of each band within the subspace of Ni $d$-orbitals, which changes with energy and momentum. (**B**) $x = 0.6$ FS in the $k_z = \pi/2$ N-P plane. Both the innermost electron pocket at the P point and the hole pocket at the N point have shrunk. (**C**) FS of the $x = 0.75$ compound in the $k_z = \pi/2$ N-P plane, close to the critical substitution where the Lifshitz transition takes place. The N pocket has all but disappeared, and the innermost P pocket continues to get smaller. Arrows indicate the coupling between the $d_{xz}$ states on the N hole pocket and on the P electron pocket (green), as well as the coupling between the $d_{yz}$ states on the same pockets (blue). (**D**) FS of the $x = 0.8$ compound in the $k_z = \pi/2$ N-P plane. The Lifshitz transition is complete, and the N hole pocket has been pushed below the Fermi level. The innermost P pocket has shrunk. Panels (E) to (H) and (I) to (L) show the energy dispersion along the P-N-P and $N - \bar{\Gamma} - N$ cuts, respectively, as a function of Sr substitution ($x = 0$, $x = 0.6$, $x = 0.75$, $x = 0.8$), showing the N-point hole pocket being pushed below the Fermi level, which is indicated by the violet dashed line.

As explained above, the top panels show the FSs in the $k_z = \pi/2$ plane; the middle panels show P-N-P energy dispersion cuts corresponding to the vertical black dashed line along the Brillouin zone edge in Fig. 4A; and the bottom panels show the $N - \bar{\Gamma} - N$ energy dispersion cuts, corresponding to the horizontal black dashed line in Fig. 4A. There is a clear qualitative change in the FS with Sr substitution, namely, a suppression of the hole pocket centered at the N point and a concomitant shrinking of the innermost P-centered electron pocket. As discussed above, such charge redistribution is expected, as the Sr atoms are nominally isovalent to the Ba atoms they are replacing, and the total charge of the system is not expected to change upon substitution. This is also consistent with our observation of the shrinking of the X-centered electron pocket for increasing Sr concentration along the $k_z = 0$ plane. As the P electron pocket shrinks, its dominant Ni $d$-orbital composition at the Fermi level also changes from $d_{x^2} - y^2$ to $d_{yz}/d_{xz}$.

Upon further increase of the chemical substitution, the N FS pocket completely disappears at the critical substitution of $x \approx 0.75$, at which point the relevant bands are fully below $E_f$. Similarly, in the view of the $k_z - k_x$ plane shown in Fig. 3 (C to E), this N FS pocket "pinches off" at the critical substitution of $x \approx 0.75$. This change in FS topology is termed a Lifshitz transition and has also been observed in connection with superconductivity in other materials such as doped $SrTiO_3$ (28, 29) and Co-doped $BaFe_2As_2$ (30). As seen from the phase diagram in Fig. 1A, the substitution level at which this Lifshitz transition occurs aligns closely with the end of the triclinic, CDW, and nematic transition lines. Moreover, as shown in (5), this is the same concentration where nematic fluctuations and the superconducting $T_c$ are enhanced. It was also reported that for Sr content greater than critical substitution, there was no long-range charge order detected in the system (5). This leads us to focus on nematicity and nematic fluctuations near critical substitution, as there is a continuous increase in superconducting $T_c$ when approaching critical substitution (from $x = 1$ to $x \sim 0.75$), concomitant with rising nematic fluctuations without any evidence of charge order.

These empirical observations thus demonstrate an unexpected link between the Lifshitz transition found in our work and the enhanced superconducting $T_c$ and nematic fluctuations observed elsewhere (see the phase diagram in Fig. 1A) (5). One possible scenario to account for this relationship borrows from results that have been widely applied to elucidate electronic nematicity in iron-based superconductors (*31*). In the itinerant model for nematicity in the iron pnictides, a hole pocket is reasonably nested with two symmetry-related electron pockets (*32*). In the coordinate system of the 1-Fe Brillouin zone, the hole pocket is centered at either Γ = (0,0,0) or M = (π,π,0), whereas the symmetry-related electron pockets are located at X = (π,0,0) and Y = (0,π,0) (see schematic of Fig. 5B). The proximity to nesting enhances density-wave fluctuations at the wave vectors corresponding to X and Y. In this case, strong nesting and a corresponding large peak in the susceptibility that would lead to long-range order is not required. The proximity to nesting between these symmetry-related pockets is what enhances the fluctuations. These are magnetic fluctuations, in the case of repulsive interactions, or charge fluctuations, in the case of attractive interactions (*31*).

Nematicity then emerges from the relative strength between the nesting-driven fluctuations at the X and at the Y wave vectors in which density-wave fluctuations around one of the ordering vectors are larger or smaller than the fluctuations around the other ordering vector. The key idea is that nematic order emerges to lift the degeneracy (i.e., "frustration") between these two fluctuation channels (*33*, *34*), which is closely related to the itinerant version of the celebrated order-by-disorder mechanism widely studied in frustrated magnets (*35*). If the nesting is good, then there will be a density-wave instability. However, as the nesting condition worsens, long-range magnetic and nematic order is suppressed toward a putative nematic QCP, where nematic fluctuations persist. In the coordinate system of the 1-Fe unit cell, the nematic order corresponds to inequivalent $x'$ and $y'$ directions, i.e., $B'_{1g}$ nematicity. Transforming back to the crystallographic 2-Fe unit, it becomes $B_{2g}$ nematicity. Several works have shown that the orbital content of the Fermi pockets plays an essential role in this scenario (*23*, *36*, *37*). For instance, random phase approximation (RPA) calculations of the five-orbital model in (*23*) showed that the largest component of the nematic susceptibility arises from the configuration of a single hole pocket (the M hole pocket in Fig. 5B) and two electron pockets (the $X/Y$ electron pockets) with similar orbital composition (in this case, $d_{xy}'$ in the figure). This theoretical framework in the iron pnictides has made experimental predictions, which appear to have been verified, such as the sign change of the resistivity anisotropy with hole doping (*38*) or the scaling between shear modulus and spin-lattice relaxation time (*39*).

We can now directly apply these results to the case of $Ba_{1-x}Sr_xNi_2As_2$. Being a weakly correlated material, as we showed here, the itinerant description should work very well. While there is no FS nesting in the $k_z = 0$ plane, there is reasonable nesting at $k_z = π/2$ between the two N hole pockets and the inner P electron pocket for a wide range of $x$ values, as shown in Fig. 4 (A and B). Similar to the iron-pnictide case described above, the nesting vectors are along the (π,0,0) and (0,π,0) directions (note, however, that the coordinate system here refers to the crystallographic 2-Ni Brillouin zone). In contrast to the pnictides, however, the orbital compositions of the hole and electron pockets do not match for low Sr concentrations. In particular, as also shown in fig. S7, for small $x$, the dominant Ni $d$-orbital for the P pocket is $d_{x^2-y^2}$, which in turn contributes almost no spectral weight to the N pockets. It is only near the Lifshitz transition that the dominant Ni $d$-orbital for the P electron pocket changes to $d_{yz}/d_{xz}$, thus matching the Ni $d$-orbital composition of the N hole pockets along the ordering vectors (blue and green arrows in Figs. 4C, 4G, and 5A). As the Sr concentration is increased, the pocket sizes become different, which suppresses long-range nematic or density-wave order toward the putative QCP. As the FS undergoes the Lifshitz transition, nesting conditions have fully deteriorated, which is consistent with the observed absence of long-range nematic or other types of order and with the subsequent enhancement of the nematic fluctuations at optimal substitution. This analysis thus potentially reveals an unexpected connection between the iron and nickel pnictides.

In summary, systematic ARPES measurements were carried out in the tetragonal phase of the $Ba_{1-x}Sr_xNi_2As_2$ system detailing the electronic structure of the Sr-substituted compounds. In addition, ab initio calculations of the electronic structure through the substitution range were performed, showing remarkable agreement with experiment. All bands have been assigned successfully, with renormalization factors close to unity, indicating the weakly correlated nature of this material system.

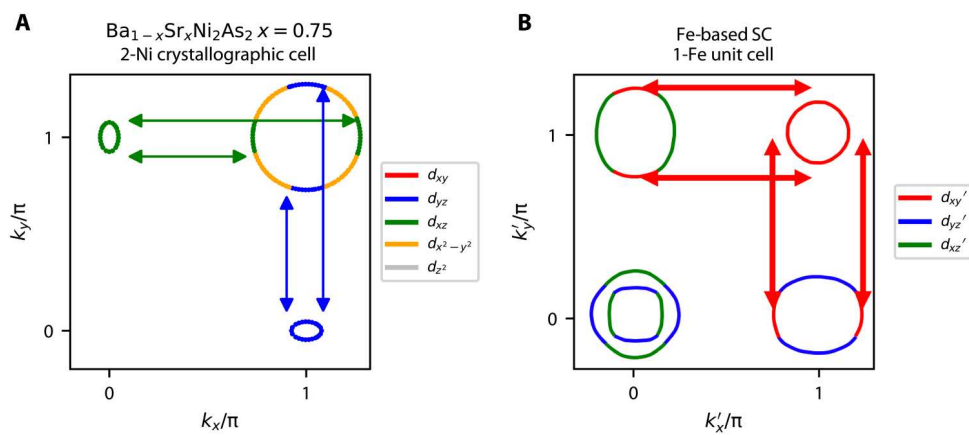

**Fig. 5. Comparison of nesting enhanced nematic fluctuations in $Ba_{1-x}Sr_xNi_2As_2$ versus the Fe-pnictides.** (**A**) $x = 0.75$ $Ba_{1-x}Sr_xNi_2As_2$ FS in the $k_z = π/2$ plane showing only the relevant N (π,0,π/2)/(0,π,π/2) hole and P (π,π,π/2) electron pockets with their dominant Ni $d$-orbital projections. Arrows delineate orbitally matched ($d_{xz}/d_{yz}$) nesting between N hole pocket and the innermost P electron pocket. In this coordinate system (2-Ni), nematic fluctuations will follow a $B_{1g}$ symmetry, which is consistent with experimental observations (*5*). (**B**) Fe-pnictide FS with Fe $d$-orbital projections from (*23*) in the 1-Fe unit cell. Arrows delineate orbitally matched ($d_{xy}'$) nesting between M (π,π,0) hole pocket and X (π,0,0)/(0,π,0) electron pockets. In this coordinate system, nematic fluctuations will follow a $B_{1g}$ symmetry. When back-folded into the crystallographic 2-Fe Brillouin zone, this yields $B_{2g}$ symmetry. Panel (B) is adapted with permission from (*23*). Copyright (2016) by the American Physical Society.

Using the ARPES-validated DFT band dispersions, we found evidence for a potential Lifshitz transition near the critical substitution $x \sim 0.75$, which onsets concomitantly with the sixfold increase of $T_c$. We used results from an itinerant model originally developed to describe nematicity of iron pnictides to argue that the approach to the Lifshitz transition sets the conditions for the emergence of enhanced electronic nematic fluctuations due to reasonably nested electron and hole pockets at the P and N points of the Brillouin zone. The nesting wave vectors connect regions of different FSs with similar $d_{yz}/d_{xz}$ dominant Ni $d$-orbital character, revealing an unexpected connection between iron and nickel pnictides. Moreover, being directed along the $x$ and $y$ axes in the crystallographic 2-Ni Brillouin zone, they support electronic nematicity in the same channel as that observed experimentally ($B_{1g}$). The scenario of enhanced nematic fluctuations enabled by the Lifshitz transition is consistent with the elastoresistance data of (5) at the $x \sim 0.75$ substitution level and with the assessment in that paper that nematicity in this compound changes from lattice-driven at small $x$ to electronically driven for larger $x$. Such nematic fluctuations are also likely responsible for the sixfold enhancement of $T_c$. While it is possible that nematic fluctuations on their own can promote pairing (14–19), a more likely scenario for Ba$_{1-x}$Sr$_x$Ni$_2$As$_2$ is that they boost a conventional electron-phonon pairing interaction (40). Future ARPES studies of the detailed temperature-dependent scattering rates on each of the major bands would likely help uncover the role of these fluctuations and their relationship to the relevant orbitals near optimal substitution. The Ba$_{1-x}$Sr$_x$Ni$_2$As$_2$ system is thus fertile ground to explore the rich physics associated with nematic quantum criticality, enabling connections with other families of superconductors (21), particularly the iron-based superconductors.

## MATERIALS AND METHODS
### Single-crystal growth
Crystals were grown from Ba, Sr, and NiAs in a ratio of 1-$x$:$x$:4, with excess NiAs to act as a flux at high temperatures. The combination was placed in an alumina crucible and sealed in a quartz tube in an argon atmosphere. The reaction mixture was heated to 1180°C, held at 1180°C for 12 hours to homogenize the mixture, and slowly cooled to 980°C at 2°C/hour. The furnace was then turned off and allowed to cool to room temperature. Crystals with dimensions of 2 mm by 2 mm by 0.5 mm, with the shortest axis universally being the $c$ axis, were extracted mechanically from the flux. The chemical compositions of the resulting crystals were determined using a combination of energy-dispersive spectroscopy and single-crystal x-ray refinements.

### Angle-resolved photoemission spectroscopy
For ARPES experiments, crystals were cleaved at pressures better than $5 \times 10^{-11}$ torr in situ and were measured at a temperature of 200 K so as to stay in the tetragonal phase. ARPES measurements shown here were carried out at beamline 5-2 of the Stanford Synchrotron Research Laboratory (SSRL) and beamline 7 (Microscopic and Electronic STRucture Observatory, MAESTRO) of the Advanced Light Source (ALS), with other additional data taken at beamline 4 (meV Resolution Line, MERLIN) of the ALS, and at beamline i05 of the Diamond Light Source. The measurements at SSRL were undertaken using a Scienta DA30 Analyzer where the energy resolution was better than 15 meV, and the angular resolution was 0.1°. The measurements at MAESTRO were taken using a modified Scienta R4000 Analyzer, where the energy resolution was ~15 meV, and the angular resolution was 0.1°. Specific experimental parameters for each spectrum and FS map are listed in the figure captions.

### Density functional theory
To compare results of ARPES measurements with theory, we calculated the electronic structure of each compound using single-unit cells with lattice parameters and Wyckoff parameters directly extracted from XRD refinements performed at 250 K. The XRD data used in this work were previously reported in (5). Substitution effects were not treated in a supercell approach, and no cell relaxation was performed. DFT calculations were carried out using the Perdew-Burke-Ernzerhof exchange correlation functional, which uses the generalized gradient approximation as implemented in the Questaal suite of electronic structure tools (41). These calculations were all-electron, full-potential calculations, in contrast to previous work that used a pseudopotential approach (22). For the $x = 0$ parent calculation, Ba was included, but for $x = 0.6$, $x = 0.75$, and $x = 0.8$ calculations, Ba was fully substituted for Sr. Calculations used a $k$-space grid of $14 \times 14 \times 14$, with a "gmax" basis set cutoff of 12. For more information about the Questaal package and its implementation of density functional methods, please refer to (41). The Fermi surface of Fig 1C was calculated using the Vienna Ab-Initio Simulatrion Package (VASP) at the generalized gradient approximation level using pseudopotentials (42, 43). The 3D rendering was performed using PyProcar (44).

## Supplementary Materials
**This PDF file includes:**
Supplementary Materials
Figs. S1 to S7
Table S1

## REFERENCES AND NOTES
1. R. M. Fernandes, A. I. Coldea, H. Ding, I. R. Fisher, P. J. Hirschfeld, G. Kotliar, Iron pnictides and chalcogenides: A new paradigm for superconductivity. *Nature* **601**, 35–44 (2022).
2. F. Ronning, N. Kurita, E. D. Bauer, B. L. Scott, T. Park, T. Klimczuk, R. Movshovich, J. D. Thompson, The first order phase transition and superconductivity in BaNi$_2$As$_2$ single crystals. *J. Phys. Condens. Matter* **20**, 342203 (2008).
3. K. Kudo, M. Takasuga, Y. Okamoto, Z. Hiroi, M. Nohara, Giant phonon softening and enhancement of superconductivity by phosphorus doping of BaNi$_2$As$_2$. *Phys. Rev. Lett.* **109**, 097002 (2012).
4. S. Lee, G. de la Peña, S. X.-L. Sun, M. Mitrano, Y. Fang, H. Jang, J.-S. Lee, C. Eckberg, D. Campbell, J. Collini, J. Paglione, F. de Groot, P. Abbamonte, Unconventional charge density wave order in the pnictide superconductor Ba(Ni$_{1-x}$Co$_x$)$_2$As$_2$. *Phys. Rev. Lett.* **122**, 147601 (2019).
5. C. Eckberg, D. J. Campbell, T. Metz, J. Collini, H. Hodovanets, T. Drye, P. Zavalij, M. H. Christensen, R. M. Fernandes, S. Lee, P. Abbamonte, J. W. Lynn, J. Paglione, Sixfold enhancement of superconductivity in a tunable electronic nematic system. *Nat. Phys.* **16**, 346–350 (2020).
6. Y. Guo, M. Klemm, J. S. Oh, Y. Xie, B.-H. Lei, L. Moreschini, C. Chen, Z. Yue, S. Gorovikov, T. Pedersen, M. Michiardi, S. Zhdanovich, A. Damascelli, J. Denlinger, M. Hashimoto, D. Lu, C. Jozwiak, A. Bostwick, E. Rotenberg, S.-K. Mo, R. G. Moore, J. Kono, R. J. Birgeneau, D. J. Singh, P. Dai, M. Yi, Spectral evidence for unidirectional charge density wave in detwinned BaNi$_2$As$_2$. *Phys. Rev. B* **108**, L081104 (2023).

7. S. Souliou, T. Lacmann, R. Heid, C. Meingast, M. Frachet, L. Paolasini, A.-A. Haghighirad, M. Merz, A. Bosak, M. Le Tacon, Soft-phonon and charge-density-wave formation in nematic $BaNi_2As_2$. *Phys. Rev. Lett.* **129**, 247602 (2022).

8. C. Meingast, A. Shukla, L. Wang, R. Heid, F. Hardy, M. Frachet, K. Willa, T. Lacmann, M. Le Tacon, M. Merz, A.-A. Haghighirad, T. Wolf, Charge density wave transitions, soft phonon, and possible electronic nematicity in $BaNi_2(As_{1−x}P_x)_2$. *Phys. Rev. B* **106**, 144507 (2022).

9. A. E. Böhmer, J.-H. Chu, S. Lederer, M. Yi, Nematicity and nematic fluctuations in iron-based superconductors. *Nat. Phys.* **18**, 1412–1419 (2022).

10. M. Merz, L. Wang, T. Wolf, P. Nagel, C. Meingast, S. Schuppler, Rotational symmetry breaking at the incommensurate charge-density-wave transition in $Ba(Ni, Co)_2(As, P)_2$: Possible nematic phase induced by charge/orbital fluctuations. *Phys. Rev. B* **104**, 184509 (2021).

11. Y. Yao, R. Willa, T. Lacmann, S.-M. Souliou, M. Frachet, K. Willa, M. Merz, F. Weber, C. Meingast, R. Heid, A.-A. Haghighirad, J. Schmalian, M. Le Tacon, An electronic nematic liquid in $BaNi_2As_2$. *Nat. Commun.* **13**, 4535 (2022).

12. S. Lee, J. Collini, S. X.-L. Sun, M. Mitrano, X. Guo, C. Eckberg, J. Paglione, E. Fradkin, P. Abbamonte, Multiple charge density waves and superconductivity nucleation at antiphase domain walls in the nematic pnictide $Ba_{1-x}Sr_xNi_2As_2$. *Phys. Rev. Lett.* **127**, 027602 (2021).

13. T. Worasaran, M. S. Ikeda, J. C. Palmstrom, J. A. W. Straquadine, S. A. Kivelson, I. R. Fisher, Nematic quantum criticality in an Fe-based superconductor revealed by strain-tuning. *Science* **372**, 973–977 (2021).

14. S. Lederer, Y. Schattner, E. Berg, S. A. Kivelson, Enhancement of superconductivity near a nematic quantum critical point. *Phys. Rev. Lett.* **114**, 097001 (2015).

15. M. A. Metlitski, D. F. Mross, S. Sachdev, T. Senthil, Cooper pairing in non-Fermi liquids. *Phys. Rev. B* **91**, 115111 (2015).

16. J. Kang, R. M. Fernandes, Superconductivity in FeSe thin films driven by the interplay between nematic fluctuations and spin-orbit coupling. *Phys. Rev. Lett.* **117**, 217003 (2016).

17. S. Lederer, Y. Schattner, E. Berg, S. A. Kivelson, Superconductivity and non-Fermi liquid behavior near a nematic quantum critical point. *Proc. Natl. Acad. Sci. U.S.A.* **114**, 4905–4910 (2017).

18. A. Klein, A. Chubukov, Superconductivity near a nematic quantum critical point: Interplay between hot and lukewarm regions. *Phys. Rev. B* **98**, 220501 (2018).

19. S. Lederer, E. Berg, E.-A. Kim, Tests of nematic-mediated superconductivity applied to $Ba_{1-x}Sr_xNi_2As_2$. *Phys. Rev. Res.* **2**, 023122 (2020).

20. A. I. Coldea, Electronic nematic states tuned by isoelectronic substitution in bulk $FeSe_{1-x}S_x$. *Front. Phys.* **8**, 528 (2021).

21. E. Fradkin, S. A. Kivelson, M. J. Lawler, J. P. Eisenstein, A. P. Mackenzie, Nematic fermi fluids in condensed matter physics. *Annu. Rev. Condens. Matter Phys.* **1**, 153–178 (2010).

22. B. Zhou, M. Xu, Y. Zhang, G. Xu, C. He, L. X. Yang, F. Chen, B. P. Xie, X. Y. Cui, M. Arita, K. Shimada, H. Namatame, M. Taniguchi, X. Dai, D. L. Feng, Electronic structure of $BaNi_2As_2$. *Phys. Rev. B* **83**, 035110 (2011).

23. M. H. Christensen, J. Kang, B. M. Andersen, R. M. Fernandes, Spin-driven nematic instability of the multiorbital Hubbard model: Application to iron-based superconductors. *Phys. Rev. B* **93**, 085136 (2016).

24. A. Fedorov, A. Yaresko, T. K. Kim, Y. Kushnirenko, E. Haubold, T. Wolf, M. Hoesch, A. Grüneis, B. Büchner, S. V. Borisenko, Effect of nematic ordering on electronic structure of FeSe. *Sci. Rep.* **6**, 36834 (2016).

25. S. V. Borisenko, D. V. Evtushinsky, Z. H. Liu, I. Morozov, R. Kappenberger, S. Wurmehl, B. Büchner, A. N. Yaresko, T. K. Kim, M. Hoesch, T. Wolf, N. D. Zhigadlo, Direct observation of spin-orbit coupling in iron-based superconductors. *Nat. Phys.* **12**, 311–317 (2016).

26. M. Yi, Y. Zhang, Z. X. Shen, D. Lu, Role of the orbital degree of freedom in iron-based superconductors. *npj Quantum Mater.* **2**, 57 (2017).

27. L. de' Medici, G. Giovannetti, M. Capone, Selective mott physics as a key to iron superconductors. *Phys. Rev. Lett.* **112**, 177001 (2014).

28. X. Lin, G. Bridoux, A. Gourgout, G. Seyfarth, S. Krämer, M. Nardone, B. Fauqué, K. Behnia, Critical doping for the onset of a two-band superconducting ground state in $SrTiO_{3−δ}$. *Phys. Rev. Lett.* **112**, 207002 (2014).

29. T. V. Trevisan, M. Schütt, R. M. Fernandes, Unconventional multiband superconductivity in bulk $SrTiO_3$ and $LaAlO_3/SrTiO_3$ interfaces. *Phys. Rev. Lett.* **121**, 127002 (2018).

30. C. Liu, T. Kondo, R. M. Fernandes, A. D. Palczewski, E. D. Mun, N. Ni, A. N. Thaler, A. Bostwick, E. Rotenberg, J. Schmalian, S. L. Bud'ko, P. C. Canfield, A. Kaminski, Evidence for a Lifshitz transition in electron-doped iron arsenic superconductors at the onset of superconductivity. *Nat. Phys.* **6**, 419–423 (2010).

31. R. M. Fernandes, A. V. Chubukov, J. Schmalian, What drives nematic order in iron-based superconductors? *Nat. Phys.* **10**, 97–104 (2014).

32. R. M. Fernandes, A. V. Chubukov, J. Knolle, I. Eremin, J. Schmalian, Preemptive nematic order, pseudogap, and orbital order in the iron pnictides. *Phys. Rev. B* **85**, 024534 (2012).

33. C. Fang, H. Yao, W.-F. Tsai, J. Hu, S. A. Kivelson, Theory of electron nematic order in LaFeAsO. *Phys. Rev. B* **77**, 224509 (2008).

34. C. Xu, M. Müller, S. Sachdev, Ising and spin orders in the iron-based superconductors. *Phys. Rev. B* **78**, 020501 (2008).

35. P. Chandra, P. Coleman, A. I. Larkin, Ising transition in frustrated Heisenberg models. *Phys. Rev. Lett.* **64**, 88–91 (1990).

36. L. Fanfarillo, A. Cortijo, B. Valenzuela, Spin-orbital interplay and topology in the nematic phase of iron pnictides. *Phys. Rev. B* **91**, 214515 (2015).

37. L. Classen, R.-Q. Xing, M. Khodas, A. V. Chubukov, Interplay between magnetism, superconductivity, and orbital order in 5-pocket model for iron-based superconductors: Parquet renormalization group study. *Phys. Rev. Lett.* **118**, 037001 (2017).

38. E. C. Blomberg, M. A. Tanatar, R. M. Fernandes, I. I. Mazin, B. Shen, H.-H. Wen, M. D. Johannes, J. Schmalian, R. Prozorov, Sign-reversal of the in-plane resistivity anisotropy in hole-doped iron pnictides. *Nat. Commun.* **4**, 1914 (2013).

39. R. M. Fernandes, A. E. Böhmer, C. Meingast, J. Schmalian, Scaling between magnetic and lattice fluctuations in iron pnictide superconductors. *Phys. Rev. Lett.* **111**, 137001 (2013).

40. A. Subedi, D. J. Singh, Density functional study of $BaNi_2As_2$: Electronic structure, phonons, and electron-phonon superconductivity. *Phys. Rev. B* **78**, 132511 (2008).

41. D. Pashov, S. Acharya, W. R. L. Lambrecht, J. Jackson, K. D. Belashchenko, A. Chantis, F. Jamet, M. van Schilfgaarde, Questaal: A package of electronic structure methods based on the linear muffin-tin orbital technique. *Comput. Phys. Commun.* **249**, 107065 (2020).

42. G. Kresse, J. Furthmü, Efficient iterative schemes for ab initio total-energy calculations using a plane-wave basis set. *Phys. Rev. B Condens. Matter* **54**, 11169–11186 (1996).

43. G. Kresse, D. Joubert, From ultrasoft pseudopotentials to the projector augmented-wave method. *Phys. Rev. B* **59**, 1758–1775 (1999).

44. U. Herath, P. Tavadze, X. He, E. Bousquet, S. Singh, F. Muñoz, A. H. Romero, PyProcar: A Python library for electronic structure pre/post-processing. *Comput. Phys. Commun.* **251**, 107080 (2020).

**Acknowledgments:** We would like to thank M. van Schilfgaarde, D. Pashov, and S. Acharya for useful discussions as well as assistance with the Questaal package. We would also like to thank A. Bostwick, C. Jozwiak, and E. Rotenberg at beamline 7 of the Advanced Light Source (ALS), T. Kim, and C. Cacho at beamline i05 of the Diamond Light Source, and J. Denlinger at beamline 4 of the ALS for their support of our experiments at each of their respective facilities. **Funding:** Work at the University of Colorado at Boulder was supported by the U.S. Department of Energy (DOE), Office of Science, Office of Basic Energy Sciences under grant no. DE-FG02-03ER46066 and the Gordon and Betty Moore Foundation's EPiQS Initiative through grant no. GBMF9458. ARPES experiments were carried out at beamline 5-2 of the SSRL. Use of the Stanford Synchrotron Radiation Lightsource, SLAC National Accelerator Laboratory, is supported by the DOE, Office of Science, Office of Basic Energy Sciences under contract no. DE-AC02-76SF00515. Additional ARPES experiments were undertaken at beamlines 4 and 7 of the ALS, which is a DOE Office of Science User Facility under contract no. DE-AC02-05CH11231, and with the support of Diamond Light Source, instrument i05 (proposal SI25827). Work at the University of Maryland was supported by the National Science Foundation grant no. DMR1905891 and the Gordon and Betty Moore Foundation's EPiQS Initiative through grant no. GBMF9071. R.M.F. (theoretical modeling) was supported by the DOE, Office of Science, Basic Energy Sciences, Materials Science and Engineering Division, under award no. DE-SC0020045. This work used the Summit supercomputer, which is supported by the National Science Foundation (awards ACI-1532235 and ACI-1532236), the University of Colorado Boulder, and Colorado State University. The Summit supercomputer is a joint effort of the University of Colorado Boulder and Colorado State University. This work also used the Alpine high-performance computing resource at the University of Colorado Boulder. Alpine is jointly funded by the University of Colorado Boulder, the University of Colorado Anschutz, and Colorado State University. **Author contributions:** D.M.N. led the ARPES experiments and analysis as well as ran most of the DFT calculations. P.H. assisted with both DFT calculations and ARPES measurements. R.K., B.S.B., and A.G.L. all assisted with ARPES experiments. C.E., P.S., J.C., and J.P. prepared the single-crystal samples and performed all of the related characterization of those samples. P.Z. performed the XRD refinement of all samples that were used in DFT calculations. M.H. and D.L. provided support of the ARPES measurements at SSRL. R.M.F. provided useful discussion and theoretical support of the conclusions and helped write the paper along with D.M.N. and D.S.D. D.S.D. directed the overall project. **Competing interests:** The authors declare that they have no competing interests. **Data and materials availability:** All data needed to evaluate the conclusions in the paper are present in the paper and/or the Supplementary Materials.

Submitted 29 April 2023
Accepted 15 September 2023
Published 18 October 2023
10.1126/sciadv.adi4966

# Science Advances

## Supplementary Materials for

**Potential Lifshitz transition at optimal substitution in nematic pnictide $Ba_{1-x}Sr_xNi_2As_2$**

Dushyant M. Narayan *et al.*

Corresponding author: Dushyant M. Narayan, duna1846@colorado.edu;
Daniel S. Dessau, dan.dessau@colorado.edu

**This PDF file includes:**

Supplementary Materials
Figs. S1 to S7
Table S1

## Supplementary Materials

**k$_z$ and photon energy selection**: To determine the correct photon energies to use for ARPES studies of Ba$_{1-x}$Sr$_x$Ni$_2$As$_2$, we undertook photon energy scans of several compounds throughout the substitution range, and compared them with their respective DFT calculations. In figure S1, we can see the resulting Fermi-surface maps in the $k_x$ - $k_z$ plane with the corresponding overlays from DFT.

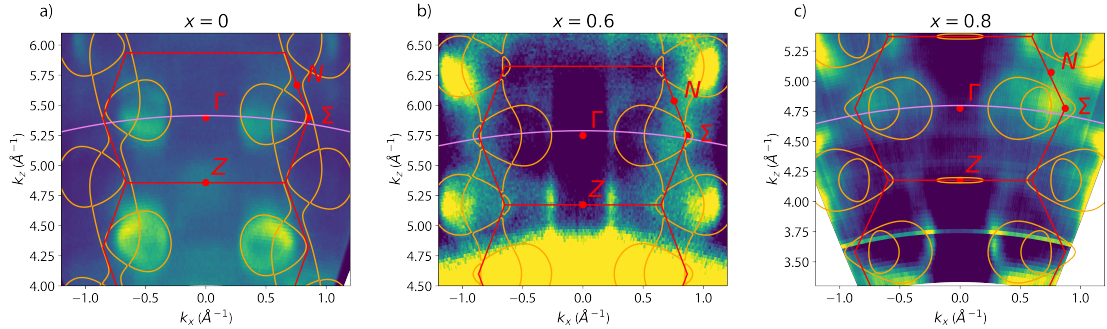

Figure S1: **Fermi-surface maps along k$_x$ − k$_z$ plane extracted from photon energy scans with DFT overlays (orange).** **a)** $x = 0$ parent compound photon energy scan taken in the triclinic phase at $T = 12$K with LH polarization from 50 to 150 eV. Inner potential was determined to be 16 eV. Violet line corresponds to 100 eV, where the rest of the data was taken for this substitution level. Overlay from DFT calculation of triclinic parent compound. **b)** $x = 0.6$ compound photon energy scan taken in the tetragonal phase at $T = 200$K with LV polarization from 65 to 160 eV. Inner potential was determined to be 16 eV. Violet line corresponds to 115 eV, where the rest of the data for this substitution level was taken. Overlay from DFT calculation of tetragonal phase using structural data extracted from XRD refinement of $x = 0.6$ crystals. **c)** $x = 0.8$ compound photon energy scan taken in the tetragonal phase at $T = 200$K with LV polarization from 30 to 120 eV. Inner potential was determined to be 17 eV. Violet line corresponds to 75 eV, where the rest of the data for this substitution level was taken. Like **b)**, overlay for **c)** also taken from tetragonal structure DFT calculation using structural data extracted from XRD refinement of $x = 0.8$ crystals.

The resulting photon energy scans were k-converted using inner potentials of 16, 16 and 17 eV for $x = 0$, $x = 0.6$ and $x = 0.8$ respectively, and show good agreement between the ARPES experimental data and DFT. Extra states can be seen in the $x = 0.6$ and $x = 0.8$ compounds that are non-dispersive and not captured by the DFT (vertical lines in the data on either side of $\Gamma$ and $Z$ points). Due to the lack of $k_z$ dispersion, which indicates confinement of the states along the real-space $z$ direction, these are likely surface states.

By performing the photon energy scans, and determining the inner potential for each compound, we can now determine the photon energies we must use to measure in-plane ARPES FS maps and cuts at the correct $k_z$ values. In figure S2, we can see the photon energies and $k_z$ values for the $\Gamma$ and $Z$ points as a function of Sr-substitution.

Using the experimental lattice parameters as extracted from XRD, we can calculate the associated reciprocal lattice vectors and distances in $k$ space between each successive $\Gamma$ and $Z$ point. Using these calculations and the inner potentials for each compound, we can create a map of the $k_z$ values vs. Sr substitution which leads to figure S2. In figure S2, we can see a smooth evolution of $k_z$ values for the high symmetry planes, and the associated photon energies needed to reach them. This is consistent with the smooth evolution of lattice parameters in the $Ba_{1-x}Sr_xNi_2As_2$ system through the substitution range. Using this data, we identify 100 eV, 115 eV, and 75 eV as photon energies to use to reach the $\Gamma$ plane in $k_z$ for the $x = 0$, $x = 0.6$ and $x = 0.8$ compounds respectively. With these photon energies in hand, we can now probe the in-plane electronic structure of $Ba_{1-x}Sr_xNi_2As_2$

**Comparison between ARPES and DFT in $Z$ plane ($k_z = \pi$)** In figure S3, we can see that in addition to the experimental spectra taken in the $k_z = 0$ plane as shown in the main text, we also took limited spectra at the $k_z = \pi$ plane. In this figure, we can see good Fermi-surface agreement between experiment and DFT for the $x = 0.6$, and $x = 0.8$ compounds. By taking this into consideration along with the good agreement in the $k_z = 0$ plane, we argue that the DFT in the $k_z = \pi/2$ plane is likely to accurately reflect the electronic structure of the real material.

**Experimental Lattice Parameters Used in DFT vs. Sr Substitution**:

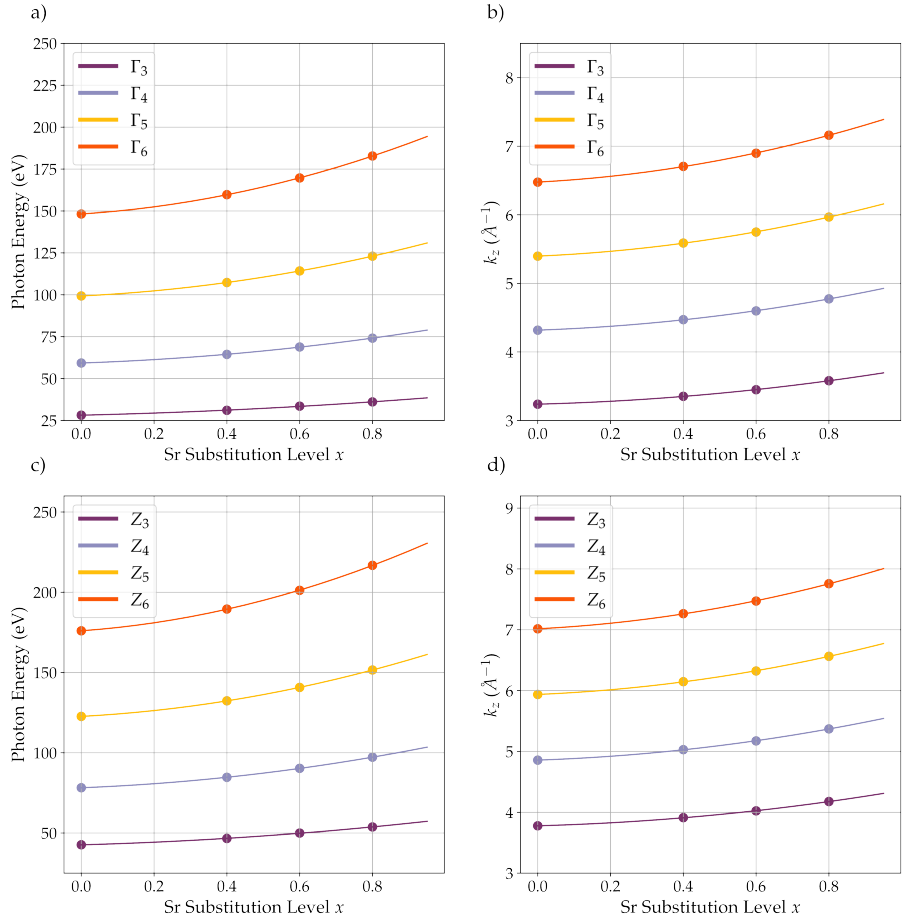

Figure S2: **$k_z$ values and the associated photon energies required to reach them as a function of Sr-substitution in the tetragonal phase.** **a**) $\Gamma$ points in photon energy as a function of Sr-substitution. Inner potentials used were 16 eV, 16 eV, 16 eV, and 17 eV for $x = 0$, $x = 0.4$, $x = 0.6$, and $x = 0.8$ compounds respectively. **b**) $k_z$ values of associated $\Gamma$ points. Points in $k$-space were calculated from experimental XRD data. **c**) $Z$ points in photon energy as a function of Sr-substitution using the same inner potentials as **a**). **d**) $k_z$ values of associated $Z$ points. Points were calculated in the same way as **b**).

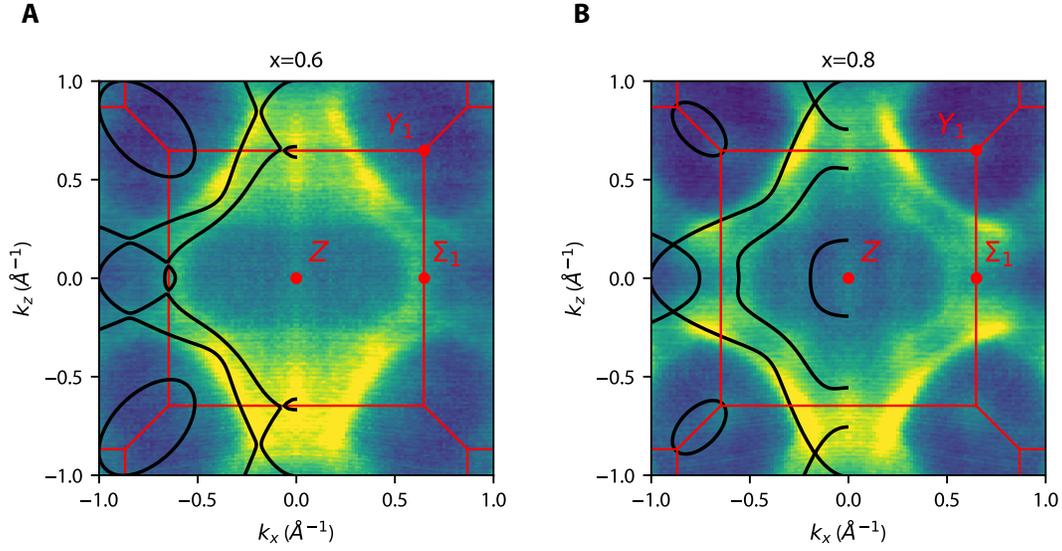

Figure S3: **ARPES Fermi-surfaces in the Z plane compared with DFT.** (**A**) $x = 0.6$ ARPES Fermi-surface taken at 91 eV, which corresponds to the $k_z = \pi$ or $Z$ plane. DFT overlaid in black. (**B**) $x = 0.8$ ARPES Fermi-surface taken at 98 eV, which corresponds to the $k_z = \pi$ or $Z$ plane. DFT overlaid in black. ARPES has been symmetrized along $k_x$ in both maps to minimize intensity variation due to matrix element effects. Both maps were taken with LH (p) polarization.

Table S1: **Experimentally determined structures used in DFT calculations.** All structures listed are the respective primitive cells at each substitution level.

| Compound | $BaNi_2As_2$ | $Ba_{0.4}Sr_{0.6}Ni_2As_2$ | $Ba_{0.25}Sr_{0.75}Ni_2As_2$ | $Ba_{0.2}Sr_{0.8}Ni_2As_2$ |
|---|---|---|---|---|
| Temperature (K) | 250K | 250K | 250K | 250K |
| Crystal System | BCT (I4/mmm) | BCT (I4/mmm) | BCT (I4/mmm) | BCT (I4/mmm) |
| Lattice Constants | a=4.1442, $\alpha = 108.53879°$ | a=4.1442, $\alpha = 109.51935°$ | a=4.1468, $\alpha = 109.84047°$ | a=4.1525, $\alpha = 110.14687°$ |
| | b=4.1442, $\beta = 108.53879°$ | b=4.1442, $\beta = 109.51935°$ | b=4.1468, $\beta = 109.84047°$ | b=4.1525, $\beta = 110.14687°$ |
| | c=6.51713, $\gamma = 90°$ | c=6.201567, $\gamma = 90°$ | c=6.108972, $\gamma = 90°$ | c=6.028115, $\gamma = 90°$ |
| Volume (Å$^3$) | 99.97472 | 93.86751 | 92.15740 | 90.77971 |
| Wyckoff Positions | Ba (1a): 0,0,0 | Sr (1a): 0,0,0 | Sr (1a): 0,0,0 | Sr (1a): 0,0,0 |
| | Ni (2d): 0.25,0.75,0.5 | Ni (2d): 0.25,0.75,0.5 | Ni (2d): 0.25,0.75,0.5 | Ni (2d): 0.25,0.75,0.5 |
| | As (2e): 0.65261,0.65261,0.30522 | As (2e): 0.64543,0.64543,0.29086 | As (2e): 0.643,0.643,0.286 | As (2e): 0.64082,0.64082,0.28164 |

In table S1 we show the experimentally refined structures used in our DFT calculations. For the parent $x = 0$ compound, Ba was used, while for the other substitution levels, Sr was fully substituted for Ba. As Sr is increased, we see an associated reduction in the cell volume, and change in the lattice constants. Interestingly, the **a** and **b** lattice constants do not shift appreciably until $x = 0.75$, where they increase sharply. For more information about the structural changes as Sr is substituted for Ba, including bond angle information, see ref. (5).

**Atomic and Orbital Projections**: In figure S4, we plot the contributions of each atom, Ba, Ni, and As, to the spectral weight on the Fermi surface of $Ba_{1-x}Sr_xNi_2As_2$. In figs. S4 a1)-a4), we plot the Fermi surface with the color at each point representing the largest contribution out of each of the three atoms. In figs. S4 b1)-b4), we resolve the angular dependence of the atomic contributions along the innermost P pocket, and in c1)-c3) as well as in d1)-d3), we do the same for the N pockets. We can see that for most of the pockets on the Fermi surface, Ni contributes the largest share of the spectral weight. For the innermost P pocket, the dominant contribution arises from As p states, but the Ni d states contribute a large fraction of the spectral weight ($\sim 45\%$). For the purposes of our analysis, we focus only on the Ni d-orbitals.

In figures S5 and S6, we see the Ni d-orbital projections of the main states along high symmetry directions. As is clearer in Fig. S6, the system shows strong hybridization, with all 5 Ni d-orbitals contributing to the spectral weight. In this orbital subspace, the main states at $E_f$ have dominant $d_{xy}$, $d_{xz}/d_{yz}$, and $d_{x^2-y^2}$ orbital characters, as shown in Fig. S5.

In figure S7, we investigate the Ni d-orbital projections on the $N$ and $P$ pockets further, where we see that $d_{xz}$, and $d_{yz}$ states become more prominent at $E_f$ as a function of Sr substitution. In figure S7 b1)-b4), the full angular-resolved Ni d-orbital composition of the innermost $P$ electron pocket is shown as a function of Sr substitution. At the substitution level close to the Lifshitz transition, the $d_{xz}$ and $d_{yz}$ components become dominant at $0°$ and $90°$, which are the directions that point towards the $N$ hole pockets. The Ni d-orbital content on the $N$ hole pockets are shown in figures S7 c1)-c3), and d1)-d3). Similarly to the $P$ pocket, the $d_{xz}$, and $d_{yz}$ orbitals become more dominant around the entire pocket as a function of Sr substitution. At the critical substitution level, the angles match with their respective counterparts on the $P$ pocket. This Ni d-orbital nesting of the states at $0°$ and $90°$ is consistent with nematic fluctuations in the $B_{1g}$ symmetry channel.

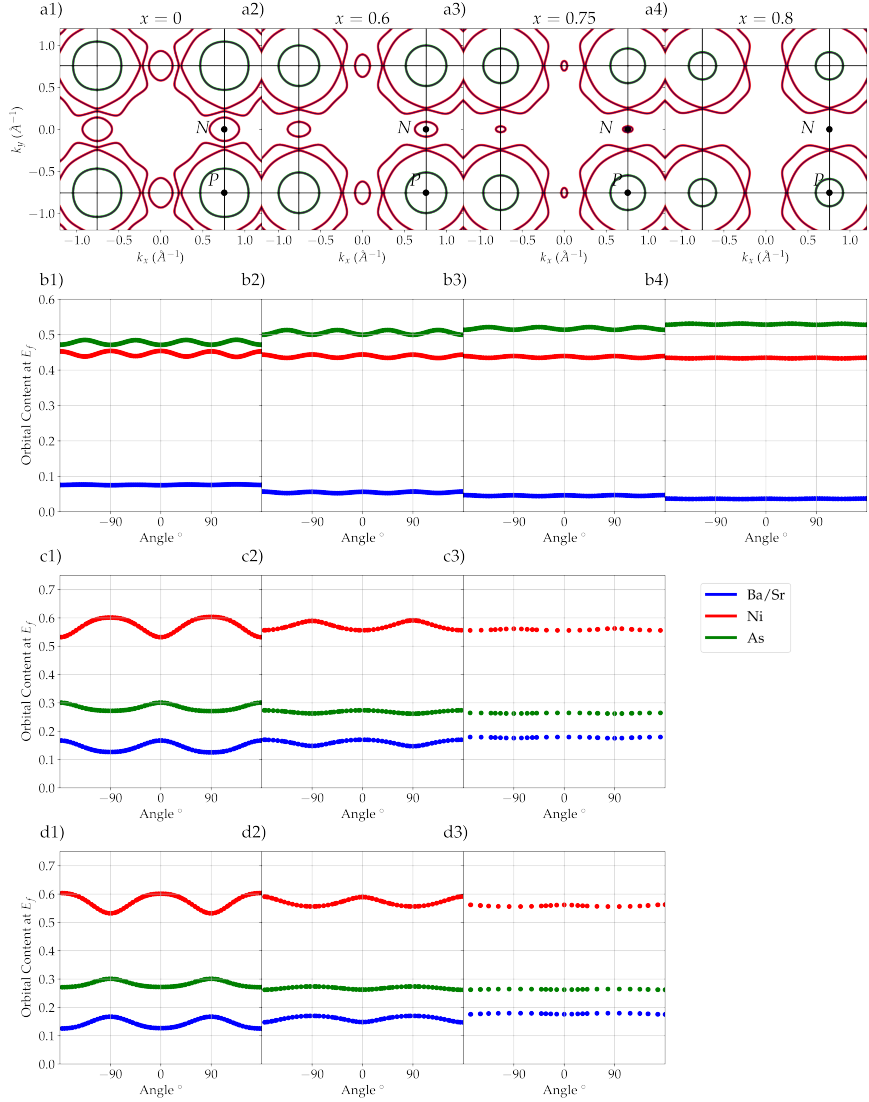

Figure S4: **Atomic projections at $E_f$.** **a1)-a4)** $N$-$P$ plane Fermi surfaces as a function of Sr substitution, showing the $N$ pocket disappearing for $x > 0.75$. Color weights at each point correspond to the atom with the largest projection at that $E$, $k$ point. **b1)-b4)** Angular dependence of the atomic projections at $E_f$ on the innermost $P$ electron pocket as a function of Sr substitution. Here, $0°$ corresponds to the $+k_x$ axis, and $90°$ corresponds to the $+k_y$ axis. **c1)-c3)** Angular dependence of the atomic projections at $E_f$ on the $N$ pocket at $k_x = 0$, $k_y = \pi$ as a function of Sr substitution. **d1)-d3)** Same as **c1)-c3)**, but for the other symmetry-related $N$ pocket found at $k_x = \pi$, $k_y = 0$.

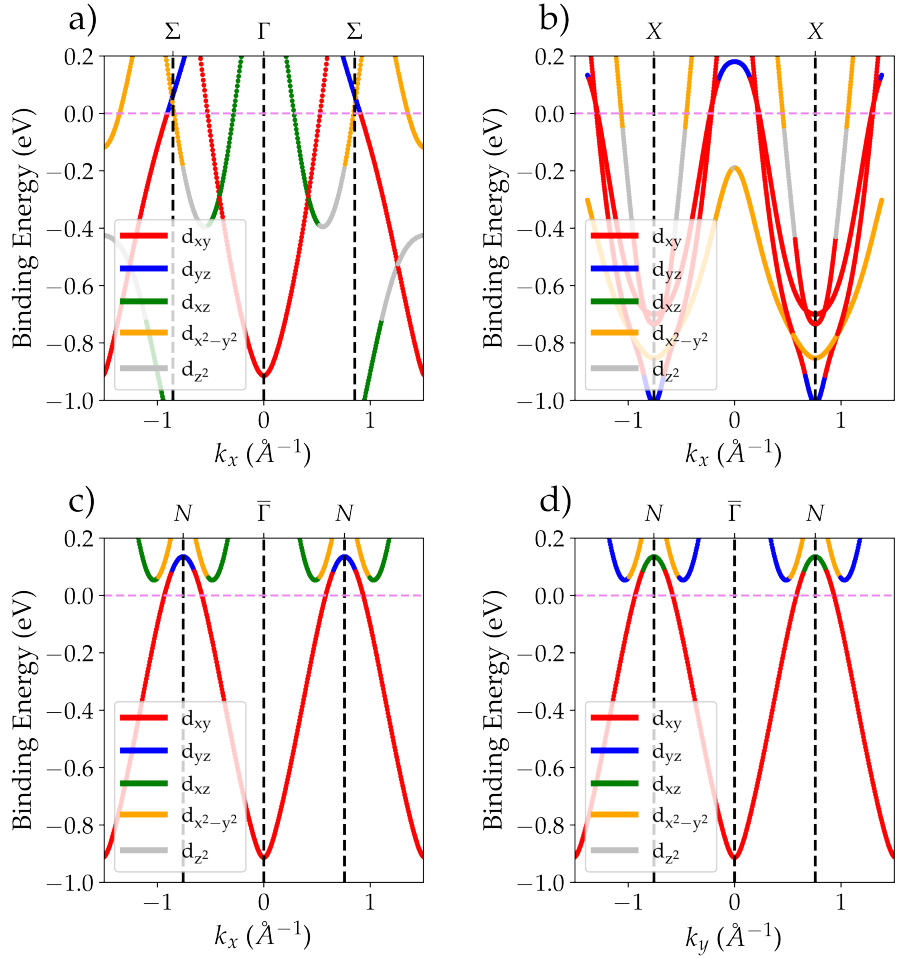

Figure S5: **E(k) dispersion for the x = 0 parent compound with Ni d-orbital projections obtained from DFT.** Black dashed lines indicate high symmetry points. The color at each band point represents the orbital with the largest spectral weight at that $E, k$ point. **a)** $\Sigma\Gamma\Sigma$ cut with orbital projection. The bands found between $\Gamma$ and $\Sigma$ are hybridized $d_{z^2}$, $d_{xz}$ and $d_{x^2-y^2}$ states, while the band centered at $\Gamma$ has majority $d_{xy}$ character. **b)** $X$-$X$ cut with orbital projection. The two outermost pockets centered at the $X$ point have dominant $d_{xy}$ orbital character, while the innermost pocket has hybridized $d_{z^2}$ and $d_{x^2-y^2}$ orbital content. **c)** $N$-$N$ Cut showing the hole pocket centered at the $N$ point. This pocket has substantial $d_{xy}$ character, but gains $d_{yz}$ character at the band-top. **d)** Same cut as **c)**, but along the $k_y$ direction, showing the greater $d_{xz}$ orbital character at the band-top.

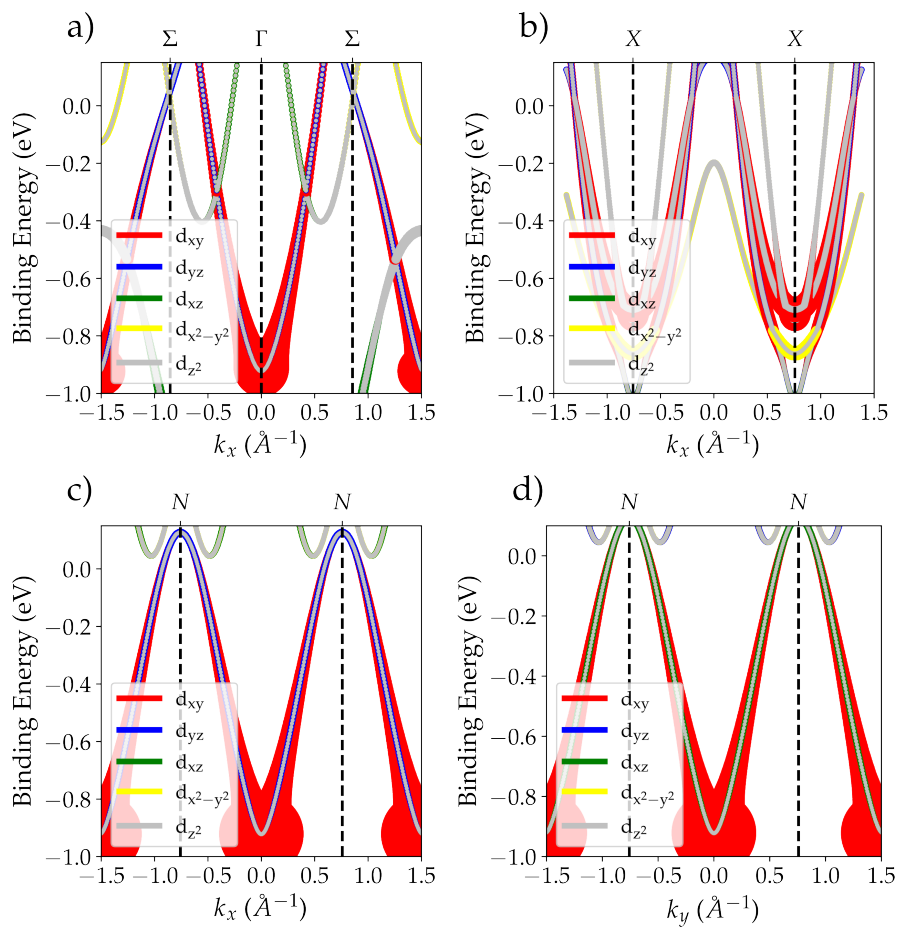

Figure S6: **E(k) dispersion for the x = 0 parent compound with Ni d-orbital projections obtained from DFT, but with the Ni d-orbital projections visualized as line thickness.** Note the strong hybridization of orbitals near $E_f$.

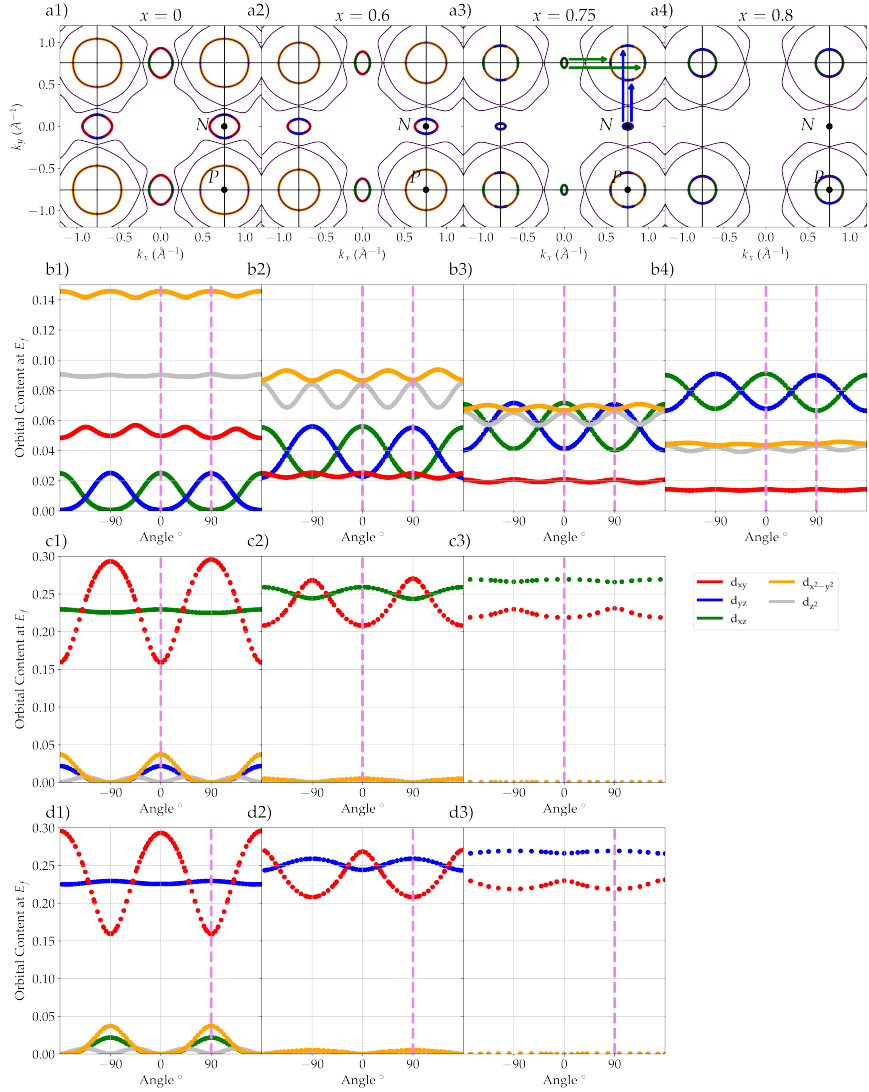

Figure S7: **Ni d-orbital projections at $E_f$. a1)-a4)** $N$-$P$ plane Fermi surfaces as a function of Sr substitution, showing the $N$ pocket disappearing for $x > 0.75$. Green and blue arrows in figure **a3)** show the $k$ vectors that connect the $d_{xz}$, and $d_{yz}$ states between the $N$ and $P$ pockets. **b1)-b4)** Angular dependence of the orbital projections at $E_f$ on the innermost $P$ electron pocket as a function of Sr substitution. Here, $0°$ corresponds to the $+k_x$ axis, and $90°$ corresponds to the $+k_y$ axis. Magenta lines correspond to the $k$ vectors in panel **a3)**. **c1)-c3)** Angular dependence of the orbital projections at $E_f$ on the $N$ pocket at $k_x = 0$, $k_y = \pi$ as a function of Sr substitution. Magenta lines correspond to the $k$ vectors connecting the $N$ and $P$ pockets along the $+k_x$ direction in panel **a3)**. **d1)-d3)** Same as **c1)-c3)**, but for the other symmetry-related $N$ pocket found at $k_x = \pi$, $k_y = 0$.